\documentclass[fleqn,twoside]{article}
\usepackage{amsmath}
\usepackage{espcrc2}
\usepackage{graphicx}
\usepackage[figuresright]{rotating}

\input{tcilatex}
\begin{document}

\title{Vortex lines in the high-T$_{c}$ superconductivity in presence of
lattice distortions.}
\author{Teresa Lenkowska-Czerwi\'{n}ska \\
%EndAName
Laboratory for Physics of Structured Media, IFTR Polish Academy of Sciences\\
\'{S}wi\,{e}tokrzyska 21, 00-049 Warsaw, Poland}

\begin{abstract}
We investigate the superconducting states of the high-T$_{c}$
superconductors which we argue to be commensurately locked by the tilted
rigid octahedral distortion. The method is based on the analysis of the
kinematics of the rigid-octahedra lattice model. The distortion modes cause
the ~competing superconducting state of the \textit{s} and $d$ type
symmetry. The vortex structure of the ~competing orders is analysed on the
ground of the Ginzburg-Landau model.

\end{abstract}

\maketitle

\section{\textbf{Introduction}}

Many low temperature properties of high-T$_{c}$ superconductivity appear to
be well described in the framework of the conventional BCS theory.
Controversions have arisen around the underdoped materials where the
instabilities to the various ~coexisting spin and charge density wave states
appear. In this regime there are no well-defined quasiparticles and no Fermi
surface in the normal state \cite{PKH}

There are several indications that the structural phase diagram in a $%
\mathrm{La_{2-\delta }M_{\delta }CuO_{4}}$ crystal (M~=~Sr, Ca, Ba, $\delta $%
-doping) involves a lower-symmetry structure at low temperatures \cite%
{Bednorz}. A tetragonal to orthorhombic distortion can be viewed as a result
of superlattice formation due to rigid octahedral tiltings. According to the
suggestions given in the paper by Jorgensen \textit{et al.} \cite{Jorgensen}%
, static displacement with octahedral tilting can be written as a
superposition of two modes with equal amplitudes, and wave vectors in
directions $[110]$ and $[-110]$. The tilting-mode distortion manifests
itself by opening up an energy gap at the Fermi surface and leads to a
structural phase transition from tetragonal $I4/mmm$ ($D_{4h}^{17}$) to
orthorhombic $Cmca$ ($D_{2h}^{18}$) phase. The structural transition takes
place within a background of superconducting order, and there is a second
order phase transitions between the phase with coexisting the ordinary
superconducting phase and lattice distortion. It is interesting to know how
the structural and superconducting phase transitions in the materials under
consideration depend on the microscopic quantities represented by the vector
of the rigid rotation of the octahedra.\newline

Many experiments and also various theoretical models suggested that the
~ground state of the underdoped superconductors $\mathrm{La_{2-\delta
}M_{\delta }CuO_{4}}$ exhibits a unidirectional density wave which consists
of a coupled spin- (SDW) and charge-density wave (CDW), called stripes. \cite%
{La214} The stripe order appears as spatial modulation of the
superconducting (SC) order parameters. On the other hand, the charge
neutrality criterion suggest a strong coupling between charge modulated
state and lattice.

\section{\textbf{Tilting Modes in the Rigid-Octahedra Lattice Model}}

We analyze a one layer of coupled rigid octahedra $\mathrm{CuO_{6}}$ in the
coordinate frame with origin in the copper central atom and axes (x,y)
directed along the diagonals of the quadratic lattice in the base layer. The
z-axis is directed perpendicularly to the plane along the c-tetragonal
direction.

The kinematic equations for the displacements of the oxygen atoms, $\vec{u}(%
\mathbf{r})$ in the $\mathrm{CuO_{2}}$ plane, are given by \thinspace Eq.$%
\left( 1\right) $ where $\vec{\varphi}$ means the octahedral rotation
vector, $\vec{d}\,_{1},\,\vec{d}\,_{2}$ are local vectors in the direction $%
(x,y)$ and $\Vert \vec{d}\,_{1}\Vert \,=\,\Vert \vec{d}\,_{2}\Vert =a$, $%
\Vert \vec{d}\,_{3}\Vert \,=\,c/2$, $a$ is the lattice constant. Because of
the connection of the neighboring octahedrals in a plane, the continuity
equation is added $\,$\ 
\begin{equation}
\begin{array}{l}
\vec{u}(\mathbf{r}\pm \frac{a}{2}\mathbf{e}_{x})\,=\,\vec{u}(\mathbf{r})\,+\,%
\frac{1}{2}\vec{d}\,_{1}\times \vec{\varphi}(\mathbf{r}) \\ 
=\,\vec{u}(\mathbf{r}\pm a\mathbf{e}_{x})\,-\,\frac{1}{2}\vec{d}\,_{1}\times 
\vec{\varphi}(\mathbf{r}\pm a\mathbf{e}_{x}),\vspace*{0.3cm} \\ 
\vec{u}(\mathbf{r}\pm \frac{a}{2}\mathbf{e}_{y})\,=\vec{u}(\mathbf{r})\,+\,%
\frac{1}{2}\vec{d}\,_{2}\times \vec{\varphi}(\mathbf{r})\, \\ 
=\,\vec{u}(\mathbf{r}\pm a\mathbf{e}_{y})\,-\,\frac{1}{2}\vec{d}\,_{2}\times 
\vec{\varphi}(\mathbf{r}\pm a\mathbf{e}_{y\,}).\vspace{0.3cm}%
\end{array}
\label{eq2}
\end{equation}%
We define the differential $D(\mathbf{r})$ and arithmetic mean values $S(%
\mathbf{r})$ lattice operators 
\begin{eqnarray*}
&&D_{1,2}\vec{u}\,\overset{\mathrm{def}}{=}\,\vec{u}(\mathbf{r}+\frac{a}{2}%
\mathbf{e}_{x,y})\,-\,\vec{u}(\mathbf{r}-\frac{a}{2}\mathbf{e}_{x,y}), \\
&&S_{1,2}\vec{u}\,\overset{\mathrm{def}}{=}\,\frac{1}{2}\,\left( \,\vec{u}(%
\mathbf{r}+\frac{a}{2}\mathbf{e}_{x,y})\,+\,\vec{u}(\mathbf{r}-\frac{a}{2}%
\mathbf{e}_{x,y})\right) .\text{ \ }
\end{eqnarray*}%
Then we can write equations (\ref{eq2}) in the form $\left(
\,\,i,j\,=\,1,2\right) $ 
\begin{equation}
\begin{array}{l}
D_{2}\,S_{1}\,\varphi _{3}\,=\,0,\,\,\,\,D_{1}\,S_{2}\,\varphi _{3}\,=\,0,%
\text{ } \\ 
D_{2}\,S_{1}\,\varphi _{2}\,+\,D_{1}\,S_{2}\,\varphi _{1}\,=\,0, \\ 
\vec{d}\,_{i}\times D_{j}\,S_{i}\,\vec{\varphi}\,=\,\vec{d}\,_{j}\times
D_{i}\,S_{j}\,\vec{\varphi}.\,\,\,%
\end{array}
\label{eq5}
\end{equation}%
We are looking for the solution of (\ref{eq5}) in the form of
sign-alternating distortion modes with reciprocal lattice vectors (\textit{%
wave vectors}) in-plane $\vec{k}=(\frac{\pi }{a},\frac{\pi }{a},0)$ for the 
\textit{plane alternating}, and $\vec{k}=(\frac{\pi }{a},0,0)$ and $\vec{k}%
=(0,\frac{\pi }{a},0)$ for the \textit{chain-alternating }modes. Because the
apex oxygen atoms above and below the CuO$_{2}$ plane are not directly
bound, we postulate the z-dependence of the mode functions. We can obtain
the three dimensional lattice model as a two-layer structure, one above the
another, at the distance $\frac{1}{2}c$, where the upper layer has an
additional shift in the plane direction above $\vec{t}_{\parallel }=(\frac{a%
}{2},\frac{a}{2})$ (because of a body centered lattice). Next we consider
the two kinds of zero modes: homogeneous - the same for all layers, and
space-alternating - changing sign under elementary translation $\vec{t}_{3}=(%
\frac{a}{2},\frac{a}{2},\frac{c}{2})$. We obtain two kinds of modes 
\footnote{%
The sign $\pm $ distinguishes the two kinds of space modes} $\,$: the
plane-alternating $\,$ $\varphi _{i\text{ }}$($i=1,2,3$), $\,$and
chain-alternating $\left( \tilde{\varphi}_{1},\tilde{\varphi}_{2}\right) $ 
\begin{equation}
\begin{array}{l}
\text{ }\varphi _{i}\,=\,q_{i}^{+}\epsilon _{+}(\mathbf{r}%
)\,+\,q_{i}^{-}\epsilon _{-}(\mathbf{r}), \\ 
\begin{array}{l}
\tilde{\varphi}_{1}\,=\,\tilde{q}_{1}^{+}\epsilon _{+}(y,z)\,+\,\tilde{q}%
_{1}^{-}\epsilon _{-}(y,z), \\ 
\tilde{\varphi}_{2}\,=\,\tilde{q}_{2}^{+}\epsilon _{+}(x,z)\,+\,\tilde{q}%
_{2}^{-}\epsilon _{-}(x,z).%
\end{array}%
\end{array}
\label{mody-z}
\end{equation}

The sign-alternating functions $\epsilon _{\pm }$ are even functions of the
arguments, and change the sign under elementary lattice translation $\mathbf{%
t}_{1}\,(a,0,0),\,\mathbf{t}_{2}\,(0,a,0),\,\mathbf{t}_{3}\,(\frac{a}{2},%
\frac{a}{2},\frac{c}{2})$: $\epsilon _{\pm }(\mathbf{r}\pm \mathbf{t}%
_{1,2})= $ $-\epsilon _{+}(\mathbf{r})$, $\varepsilon _{\pm }(\mathbf{r}\pm 
\mathbf{t}_{3})=\pm \varepsilon _{\pm }(\mathbf{r}).$

The amplitudes $q_{i}^{\pm },\tilde{q}_{1,2}^{\pm }$ of the functions $%
\varphi _{i},\,\tilde{\varphi}_{1,2}\,$ (i=1,2,3) are the micro-order
parameters. The macro-order parameters are constructed from the
microparameters as translationally invariant polynomials of the degrees $%
n\leq 4$. Macroparameters change under the action of the point group
elements. In dealing with the symmetry class $D_{4h}$, for the generating
elements of the group we choose the rotations $C_{4}^{1}$, inversion $I$ and
reflection $\sigma _{d}$. Finally, the translational and rotational
invariant polynomials for $D_{4h}^{17}$ - the macro-order parameters are
found \cite{TLC}. The distortion instability can change the lattice symmetry
due to coupling between the distortion and lattice deformation \cite{TLC-DR}%
. These allow one to construct the appropriate thermodynamic potential for
describing the second order structural phase transitions in the
superconducting materials of the La214 class. Some invariants could lead to
phase transitions in low symmetry phase - orthorhombic $D_{2h}$ and
monoclinic $C_{2}$, and to the phase transitions without the symmetry change %
\cite{TLC}.

\section{\textbf{Symmetries of the Cooper Pair Function}}

In a superconducting material $\mathrm{La_{2-\delta }Sr}_{\mathrm{\delta }}%
\mathrm{CuO_{4}}$ two different superconducting phases, in orthorombic and
tetragonal material structures have been distinguished \ \cite{Keimer}. The
temperature of the both structural and superconducting phase transitions
depends on the doping. The critical temperature for superconducting phase
transition, $T_{s}$, has a maximum very near the intersection point of the
structural and superconducting phase lines in the ($\delta ,T$) plane
(optimal doping). In the neighborhood of this point, the structural and
superconducting order parameters are strongly correlated and the order
parameter of the lattice distortion allows a rich variety of phases and
phase transitions in the presence of the background superconducting order.

The finite distortion produces a confining potential for the charge carries
and we thereby establish a direct relation between the symmetry of the
structural and superconducting order parameters. Thus, about the
superconducting state under consideration we assume: $\left( i\right) $ The
singlet superconducting state which means, that the resulting spin of the
electron pairs (Cooper pairs) equal zero. Therefore, the symmetry group $%
\mathcal{G}$ under consideration contains the space group $G=D_{4h}^{17}$,
time inversion $\mathcal{K}$ and gauge $U(1)$ groups, $\mathcal{G}=G\times 
\mathcal{K}\times U(1)$. $\left( ii\right) $ The wave function of the Cooper
pairs in the form $\Psi (\mathbf{r}_{1},\mathbf{r}_{2})\,=<c_{\uparrow }(%
\mathbf{r}_{1})c_{\downarrow }(\mathbf{r}_{2})>=\psi (\mathbf{r}%
_{c})\vartheta (\mathbf{r}_{1}-\mathbf{r}_{2}),$ where $\mathbf{r}_{c}$
stands for the center of mass of the two electrons located in the lattice
points ($\mathbf{r}_{1},\mathbf{r}_{2}).\,$The function $\psi (\mathbf{r})$
changes under elementary lattice translations. The function $\vartheta (%
\mathbf{r}_{1}-\mathbf{r}_{2})$ depends on the rotational symmetry of the
Cooper pair. $\left( iii\right) $The symmetries of the superconducting and
distortions order parameters are conformable. We therefore postulate: $\psi (%
\mathbf{r})=$ $\varepsilon ^{\pm }(\mathbf{r})$ for the plane alternating
modes, and $\psi (\mathbf{r})=\{\varepsilon ^{\pm }(y,z),\,\varepsilon ^{\pm
}(x,z)\}$ for the chain alternating modes.

The superconducting order parameter (or Cooper pair wave function) involve
relative and a centre of mass pieces, and the latter is what enters the
standard Ginzburg-Landau theory. The general form of Landau free energy has
been discussed in the paper \cite{TLC}

\section{\textbf{Vortices in the inhomogeneous superconducting states}}

Now we focus our attention on the state which are chain-alternating in the $%
\left( x,y\right) $ copper-oxide plane, and uniform in the space direction $%
z $ ($\mathbf{k=}$\ $\mathbf{k}_{x}+\mathbf{k}_{y})$ 
\begin{eqnarray*}
\psi _{x}\text{ }\left( \mathbf{k}\right) \text{{}} &=&\text{{}}\epsilon
_{+}(x)\eta _{x},\text{ \ \ }\psi _{y}\left( \mathbf{k}\right) \text{ =}%
\epsilon _{+}(y)\eta _{y}\text{, \ \ \ } \\
\epsilon _{+}(x) &=&e^{i\mathbf{k\cdot x}},\text{ \ \ \ \ }\epsilon
_{+}(y)=e^{i\mathbf{k\cdot y}}.
\end{eqnarray*}%
In the absence of the external magnetic field, the complex \ functions $\eta
_{x,y}$ are space independent.

We define the order parameters for two nondegenerate states \footnote{%
The degeneration is removed by the lattice distortion.} 
\begin{eqnarray}
\psi _{s}\left( \mathbf{k}_{+}\right) &=&\psi _{x}\left( \mathbf{k}\right)
+\psi _{y}\left( \mathbf{k}\right) \text{,} \\
\text{ \ \ \ }\psi _{d}\left( \mathbf{k}_{-}\right) &=&\psi _{x}\left( 
\mathbf{k}\right) -\psi _{y}\left( \mathbf{k}\right) .\text{ }
\end{eqnarray}%
After little algebra we can see that $\psi _{s}$ has the full $s$-type
rotational symmetry while $\psi _{d}$ $\left( \text{the lower-energy mode}%
\right) $ is $d_{x^{2}-y^{2}}$ - type (change the sign in the diagonal
direction). \ In the external magnetic field directed ~perpendicularly to
the superconducting plane, the amplitudes $\psi _{s}$ and $\psi _{d}$ become
the space-dependent order parameters while the ~functions $\zeta $ remain
field independent .

Thus under conditions \textbf{k}$_{+}$=-\textbf{k}$_{-}$ the simplest free
energy density allowed by the symmetry results from the coupling between the
distortion-dependent local order parameters 
\begin{eqnarray*}
F &=&\alpha _{s}\left| \psi _{s}\right| ^{2}+\alpha _{d}\left| \psi
_{d}\right| ^{2}+\beta _{s}\left| \psi _{s}\right| ^{4}+\beta _{d}\left|
\psi _{d}\right| ^{4}+ \\
&&\beta _{m}\left( \psi _{s}^{\ast 2}\psi _{d}^{2}+\psi _{s}^{2}\psi
_{d}^{\ast 2}\right) +\gamma _{s}\left| \mathbf{D}\psi _{s}\right| ^{2}+ \\
&&\gamma _{d}\left| \mathbf{D}\psi _{d}\right| ^{2}+\gamma \lbrack \left(
D_{x}\psi _{s}\right) ^{\ast }\left( D_{x}\psi _{d}\right) - \\
&&\left( D_{y}\psi _{s}\right) ^{\ast }\left( D_{y}\psi _{d}\right) ]+\frac{%
h^{2}}{8\pi }.
\end{eqnarray*}

Here $\mathbf{D=}$ $-i\mathbf{\nabla }-e^{\ast }\mathbf{A/}\left( h\text{c}%
\right) $, $\alpha _{s\left( d\right) }=a_{\delta }\left( T-T_{s\left(
d\right) }\right) $, $a_{\delta }$ depends on the doping, $\delta $ \cite%
{Keimer} and $\mathbf{A}$ means the vector potential of the external field.
Following the standard procedure \cite{TLC} \cite{TLC-DR} we derive the
system of nonlinear, coupled equations for the order parameters by ~varying
the free energy with respect to the field $\psi _{d}^{\ast }$\ and $\psi
_{s}^{\ast }$ 
\begin{equation}
\begin{array}{l}
\left( \gamma _{d}D^{2}\right) \psi _{d}+\gamma \left(
D_{x}^{2}-D_{y}^{2}\right) \psi _{s}+2\beta _{d}\left| \psi _{d}\right|
^{2}\psi _{d} \\ 
+2\beta _{m}\psi _{s}{}^{2}\psi _{d}^{\ast }+\alpha _{d}\psi _{d}=0, \\ 
\left( \gamma _{s}D^{2}\right) \psi _{s}+\gamma \left(
D_{x}^{2}-D_{y}^{2}\right) \psi _{d}+2\beta _{d}\left| \psi _{s}\right|
^{2}\psi _{s} \\ 
+2\beta _{m}\psi _{d}^{2}\psi _{s}{}^{\ast }+\alpha _{s}\psi _{s}=0.%
\end{array}%
\end{equation}

Like in \cite{vortex in d-wave} the asymptotic solutions of the equations
(inside and outside of the vortex core) can be found for the isolated vortex
line in the strongly II-type materials . They allow us to conclude that near
the structural phase transitions caused by distortion modes the d-wave
superconducting vortices ~arise close to $T_{d}$ where $\left| \psi
_{s}\right| \ll \left| \psi _{d}\right| .$ The ~competing orders of the
s-wave reach a maximum inside the core of vortices and decay algebraically
outside the core region (where the d-wave component reach the maximum bulk
value). As a result, in the presence of a vortex line, the four s-wave
vortices appear outside the vortex core due to the collinear distortion
modes \cite{d-wave}.

\begin{center}
\textbf{4. Conclusion}
\end{center}

It has been shown on the ground of kinematics lattice model \cite{TLC} that
the chain-alternating superconducting states may appear in the lattice under
the tilt-mode distortions. From the point of view of the rigid octahedral
lattice as a micropolar Cosserat model \cite{Rymarz} the structural
microscopic order parameters for the tilting, sign-alternated modes for one $%
\mathrm{CuO_{2}}$ base layer has been defined. The macroscopic properties
are determined by the quantities invariant under any microscopic
translation. Therefore, the macro-order parameters were constructed from the
translation invariants of the microparameters. The macro-order parameters,
which are zero above the critical temperature $T_{c}$ of the second order
phase transition and nonzero below $T_{c}$, allowed us to distinguish among
the various symmetries of the crystalline states. We thereby establish a
direct relation between the distortion modulations and the superconducting
state. The appropriate symmetries of the superconducting order parameters
have been discussed. We therefore conclude that the superconducting states
are closely related to the distortion modulation of the rigid CuO$_{6}$
octahedra.

The finite distortion produces a confining potential for charge carries and
causes the distribution of the charge (CDW) and Cooper pair (SC) density in
the CuO$_{2}$ plane to be inhomogeneous. From the symmetry considerations we
conclude that the phase modulations of $s$ and $d$-wave Cooper pairs arise
in the presence of the isolated vortex line.

\bigskip

I am grateful to Prof. Dominik Rogula for many valuable discussions and
useful suggestions.\newline

The Scientific Research Committee (Poland) under grant No. 5 T0 7A 040 22
~supported this work.

\bigskip

\bigskip


\begin{thebibliography}{99}
\bibitem{PKH} L.P. Pryadko, S. Kivelson, D.W. Hone, \textit{\ }Phys. rev.
Lett, \textbf{80} (1998) 5651.

\bibitem{Bednorz} J. G. Bednorz and K. A. M\"{u}ller, Rev. Mod. Phys. 
\textbf{60} (1988) 585.

\bibitem{Jorgensen} J. D. Jorgensen, H.-B. Sch\"{u}ttler, D. G. Hinks, D. W.
Capone, II, K. Zhang, M. B. Brodsky, and D. J. Scalapino, Phys. Rev. Lett., 
\textbf{58} (1987) 1024.

\bibitem{TLC} T. Lenkowska-Czerwi\'{n}ska, J. Tech. Phys., \textbf{40},
(1999) 3.

\bibitem{TLC-DR} D. Rogula, and T. Lenkowska-Czerwi\'{n}ska, J. Tech. Phys., 
\textbf{42 }(2001) 269.

\bibitem{Rymarz} Czes\l aw Rymarz, Mechanics of Continua (in Polish), PWN
Warszawa, 1993.

\bibitem{Keimer} B. Keimer, N. Belk, R. J. Birgeneau, A. Cassanho, C. Y.
Chen, M. Greven, M. A. Kastner, A. Aharony, Y. Endoh, R. W. Erwin, and G.
Shirane, Phys. Rev. B \textbf{46}, (1992)14034-14053.

\bibitem{vortex in d-wave} Y. Ren, J. H. Xu, and C. S. Ting, Phys. Rev.
Lett. \textbf{74 }(1995) 3680.

\bibitem{La214} C. Panagopoulos, J. R. Cooper, T. Xiang, Y. S. Wang, and C.
W. Chu, Phys. Rev. B 61, R3808-R3810 (2000); Y. S. Lee, R. J. Birgeneau, M.
A. Kastner, Y. Endoh, S. Wakimoto, K. Yamada, R. W. Erwin, S.-H. Lee, and G.
Shirane, Phys. Rev. B 60, 3643-3654 (1999); J.~Orenstein and A.~Millis,
Science \textbf{288} (2000) 468.

\bibitem{d-wave} M. Franz, C. Kallin, P. I. Soininen, A. J. Berlinsky, and
A. L. Fetter, Phys. Rev. B \textbf{53}, (1996) 5795-5814.
\end{thebibliography}
\end{document}